# Compact broadband thermal absorbers based on plasmonic fractal metasurfaces

ROMIL AUDHKHASI,[1,†] VIRAT TARA,[1,†] RAYMOND YU,[2] MICHELLE L. POVINELLI[2] AND ARKA MAJUMDAR[1,3,*]

[1]Department of Electrical and Computer Engineering, University of Washington, Seattle, USA
[2]Department of Electrical and Computer Engineering, University of Southern California, Los Angeles, USA
[3]Department of Physics, University of Washington, Seattle, USA
[†]These authors contributed equally
[*]arka@uw.edu

**The ability to efficiently absorb thermal radiation within a small material volume is crucial for the realization of compact and high-spatial resolution thermal imagers. Here we propose and experimentally demonstrate a compact plasmonic metasurface for broadband absorption in the 6 – 14 µm wavelength range. As opposed to previous works, our metasurface leverages strongly localized electromagnetic modes to achieve high absorption within a compact form factor. We numerically investigate the spectral response of finite arrays of fractals and show that the absorption enhancement provided by arrays with greater than 6x6 fractals covering a total area of only 30x30 µm² is similar to that of an infinitely periodic array. Furthermore, we experimentally validate our metasurface's absorption enhancement and demonstrate a good qualitative agreement between the measured and simulated spectral responses. Owing to its ability to achieve broadband absorption enhancement in a compact footprint, our metasurface provides new avenues for the realization of next-generation infrared sensors and bolometers.**

## 1. INTRODUCTION

The ability to detect thermal radiation across a broad spectral range is crucial for a wide variety of applications in defense and surveillance [1], medicine [2] and agriculture [3]. The field of thermal sensing has traditionally been dominated by two technologies: photodetectors and bolometers. Photodetectors, which rely on photon absorption to generate electron-hole pairs are fundamentally limited by the bandgap of the photosensitive material and the requirement of cryogenic temperatures to operate in the long-wave infrared (LWIR) range [4]. Recent approaches to overcome these limitations by leveraging two-dimensional (2D) materials for strong light-matter interaction face challenges with large-area fabrication [5, 6]. In contrast, bolometers, which utilize thermally absorbing materials to convert an increase in temperature to a change in electrical resistance, are a promising alternative for thermal sensing. Ongoing efforts to develop highly miniaturized micro-bolometer arrays depend on approaches to achieve high thermal sensitivity while minimizing pixel crosstalk. These, in turn, necessitate strategies to achieve strong thermal absorption within a compact form factor.

Recent advances in photonic metamaterials have provided new avenues for the realization of highly customized thermal emitters [7, 8] and absorbers [9-11]. Such metasurfaces use an array of optical resonators to excite electromagnetic (EM) modes within absorptive materials to produce a predefined spectral response. While recent studies have adapted this approach to achieve broadband absorption within a subwavelength material thickness, their reliance on delocalized EM modes necessitates a large lateral span for the absorber [12-16]. Utilization of such absorbers for developing compact microbolometers can impose limitations on the minimum achievable pixel size (and hence the spatial resolution) [17] as well as the operation speed. Therefore, compact photonic absorbers leveraging localized EM modes are critical for the realization of fast, high-spatial resolution microbolometer arrays.

Here we propose and experimentally demonstrate a compact plasmonic absorber for broadband operation in the 6 – 14 µm wavelength range. Our metasurface leverages strongly localized EM modes supported by platinum fractal resonators in conjunction with the inherent material loss of non-stoichiometric silicon nitride ($SiN_x$) to achieve broadband absorption within a small volume. We numerically investigate the absorption enhancement provided by finite arrays of fractals and show that a 6x6 fractal array covering a total area of only 30x30 µm² absorbs as much as an infinitely periodic array. Furthermore, we fabricate and experimentally characterize fractal arrays with varying sizes and show a good qualitative match with our numerical predictions.

Our metasurface provides new avenues for the development of fast, high-resolution thermal imaging systems. Additionally, this study highlights the merits of using localized photonic resonators for developing compact, multi-functional metasurfaces across different application areas.

## 2. RESULTS AND DISCUSSION

Our absorber consists of an 800 nm thick layer of non-stoichiometric silicon nitride ($SiN_x$) sandwiched between a two-dimensionally periodic array of 3rd generation Gosper fractals [18, 19] made of platinum (Pt) on the top and a Pt back reflector at the bottom. The period of the fractal array is 5 μm, the thickness of the individual fractals is 60 nm while that of the Pt back reflector is 300 nm. A unit cell of the structure is presented in Fig. 1(a). We determine the optical response of the structure by simulating it in Lumerical FDTD Solutions. Periodic boundary conditions are used along the faces of the unit cell perpendicular to the x and y axes while perfectly matched layer (PML) boundary conditions are used along those perpendicular to the z axis. The optical constants of $SiN_x$ used in the metasurface are determined from ellipsometry measurements (see supporting information for details) while those of Pt are taken from Ref. [20]. The structure is illuminated with a normally incident plane wave from the top and reflection is recorded in the 6 – 14 μm wavelength range using a power flux monitor. As there is no transmission through the structure due to the presence of 300 nm thick Pt back reflector, the absorption is calculated as (1 – reflection).

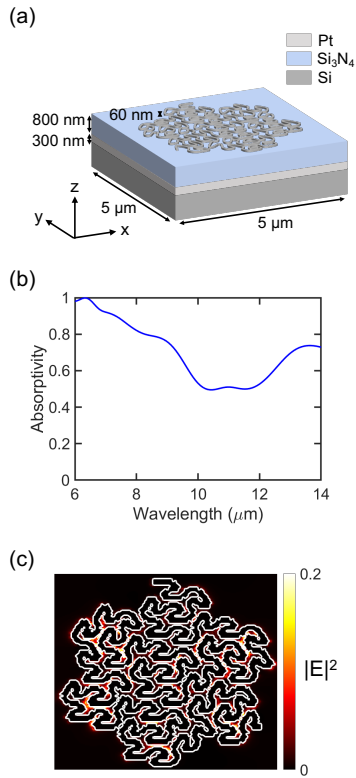

Fig. 1. (a) Schematic of a unit cell of the absorber. (b) Numerically calculated absorption spectrum of the metasurface in the 6 – 14 μm wavelength range. (c) Spatial profile of the electric field intensity within a single unit cell of the metasurface at a wavelength of 6.32 μm.

Figure 1(b) presents the numerically-calculated absorption spectrum of the structure. We observe that the metasurface exhibits broadband absorption over the whole 6 – 14 μm wavelength range with an absorptivity close to 1 at a wavelength of 6.32 μm. The percentage of the total incident power absorbed by the metasurface is greater than 50% at all wavelengths, while the spectrally averaged absorption is 71%. These results provide strong evidence for the ability of our metasurface to serve as a broadband absorber over a spectral bandwidth of 8 μm. To understand the physical basis of the absorption behavior of our metasurface, we study the electric field intensity profile within a single unit cell at a wavelength of 6.32 μm (Fig. 1(c)). For ease of visualization, we outline the fractal structure with a white line. We observe that the field is strongly localized within the gaps between the various metallic segments forming the fractal with nearly zero field leakage to other regions of the unit cell. This is a direct consequence of the localized nature of the plasmonic modes supported by the metasurface. We note that while the fractal array contributes to the high absorption observed below a wavelength of 10 μm, the increase in absorption between 12 and 14 μm is a result of the intrinsic material loss of the silicon nitride layer of the metasurface.

We note that the localized EM modes supported by our metasurface allow for the possibility of achieving broadband absorption enhancement in finite arrays of fractals. This is in contrast to a vast majority of previous works that have to rely on very large metasurface areas owing to their dependence on delocalized modes. To illustrate this point, we numerically calculate the spectral response of an $N$x$N$ array of fractals for different values of $N$. The center-to-center separation between the fractals is kept fixed at 5 μm, similar to Fig. 1. The finite fractal arrays are simulated with a Gaussian beam source with a spot size of 20 μm to mimic the source available in our metasurface characterization setup. A fixed lateral simulation domain size of 45 μm is used for all array sizes with the $SiN_x$ spacer and Pt back reflector extending through the edges of the simulation domain. Perfectly matched layer boundary conditions are used along the edges of the domain perpendicular to the *x* and *y* directions in addition to those perpendicular to the *z* direction.

Figure 2(a) provides the simulated spectral response of an $N$x$N$ fractal array for a few different values of $N$. For comparison, we also present the absorption spectrum of a periodic array of fractals (black dotted line, same as Fig. 1(b)). We observe that a 2x2 fractal array exhibits broadband absorption with absorptivity values greater than 0.4 for all wavelengths and a peak absorptivity of 0.77 at a wavelength of 13.12 μm. Additionally, the spectral response of the 8x8 fractal array matches closely with that of the periodic array, thus validating the localized nature of the absorption enhancement exhibited by our metasurface. We note that the absorptivity values for wavelengths smaller than 10 μm increase with the size of the fractal array while those for wavelengths larger than 12 μm exhibit a slight decrease. This can be attributed to the fact that an increased fractal array size causes a reduction in the area of the silicon nitride layer exposed to incident radiation, thereby reducing its contribution to the overall absorption.

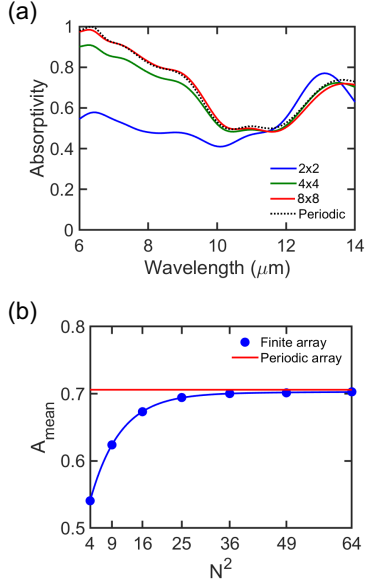

Fig. 2. (a) Simulated absorption spectra for an $N$x$N$ fractal array for a few different values of $N$. The spectrum for an infinitely periodic array is shown as a dotted black line for comparison. (b) Dependence of spectrally averaged absorption of an $N$x$N$ fractal array on $N^2$. The value of $A_{mean}$ for a periodic array is shown by the solid red line.

Next, we determine the smallest fractal array size that can provide absorption enhancement similar to an infinitely periodic array. Figure 2(c) presents the dependence of spectrally averaged absorptivity $A_{mean}$ of a finite fractal array on the number of fractals $N^2$. For comparison, the spectrally averaged absorption of a periodic array of fractals (corresponding to the dotted black line in Fig. 2(a)) is shown as a solid red line. We observe that $A_{mean}$ increases double-exponentially with $N^2$ and saturates close to the red line for arrays with 6x6 fractals or more.

To experimentally test the absorption performance of our proposed metasurface, we fabricate fractal arrays with different sizes. The left and right panels of Fig. 3(a) show the scanning electron microscope (SEM) image of a 2x2 and 8x8 array, respectively. Figure 3(b) presents the experimentally determined absorption spectra of the fabricated metasurfaces. The spectrum for a given array size is obtained by measuring its reflection spectrum using a Fourier Transform Infrared (FTIR) spectrometer and subtracting it from 1. Details of the fabrication procedure and FTIR measurements are provided in the supporting information document. The measured spectra show similar features as the calculated spectra presented in Fig. 2(a), with an expected increase in absorptivity values for larger array sizes.

To better understand the dependence of achieved absorption on the number of fractals, we present the dependence of $A_{mean}$ on $N^2$ in Fig. 3(c). For comparison, we also present the simulated $A_{mean}$ values (same as Fig. 2(b)) for different array sizes. We observe an expected increase in the measured $A_{mean}$ with $N^2$, with the increase for $N^2 = 9$ falling below the trendline potentially due to fabrication imperfections. We note that the measured $A_{mean}$, values show a steeper increase with $N^2$ as compared to the theoretical values. This is potentially a consequence of fabrication imperfections such as sidewall roughness which may contribute to an increase in inter-fractal coupling, thereby resulting in a greater dependence of the absorption enhancement on the number of fractals. Despite this discrepancy, the difference between the measured and simulated $A_{mean}$ values reduces for larger array sizes, with the values matching closely for an 8x8 fractal.

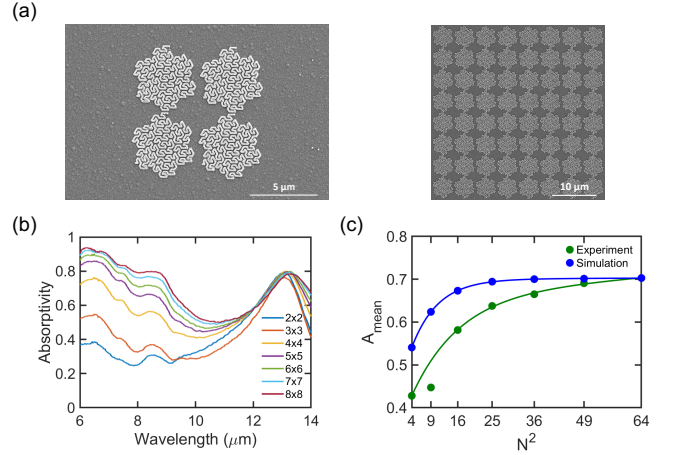

Fig. 3. (a) SEM images of a 2x2 (left panel) and 8x8 (right panel) fractal array. (b) Measured absorption spectra for an $N$x$N$ fractal array for a few different values of $N$. (c) Dependence of measured spectrally averaged absorption of an $N$x$N$ fractal array on $N^2$. The theoretical values of $A_{mean}$ for different fractal array sizes are shown by the blue dots.

## 3. CONCLUSION

We proposed and experimentally demonstrated a plasmonic fractal metasurface for broadband absorption in the 6 – 14 μm wavelength range. The absorption enhancement is enabled by leveraging strongly localized plasmonic modes supported by Pt fractal resonators in conjunction with a cavity formed from a lossy $SiN_x$ layer. Owing to the localized nature of the plasmon modes, our metasurface is able to absorb incoming radiation within a compact form factor as opposed to a vast majority of previous works. We numerically investigated metasurfaces consisting of a finite number of fractals and showed that a 6x6 fractal array is sufficient to achieve absorption similar to that of an infinitely periodic array. Additionally, we validated the ability of our metasurface to achieve high absorption within a compact form factor by fabricating and experimentally characterizing finite arrays of fractals. We believe that the proposed metasurface can provide new avenues for the development of highly sensitive infrared photodetectors and microbolometers. More importantly, photonic resonators supporting highly localized EM modes such as the ones discussed in this study could pave the way for the next generation of compact, multi-functional photonic metasurfaces.

**Funding.** The program is funded by ONR. Part of this work was conducted at the Washington Nanofabrication Facility / Molecular Analysis Facility, a National Nanotechnology Coordinated Infrastructure (NNCI) site at the University of Washington with partial support from the National Science Foundation via awards NNCI-1542101 and NNCI-2025489.

**Acknowledgment.** The authors would like to thank Fengnian Xia and Mo Li for helpful discussions.

**Disclosures.** The authors declare no conflicts of interest.

**Data availability.** Data underlying the results presented in this paper are not publicly available at this time but may be obtained from the authors upon reasonable request.

**Supplemental Document.** See Supplement 1 for supporting content.

# Supporting information for "Compact broadband thermal absorbers based on plasmonic fractal metasurfaces"

Romil Audhkhasi[1,†], Virat Tara[1,†], Raymond Yu[2], Michelle L. Povinelli[2] and Arka Majumdar[1,3,*]

[1]Department of Electrical and Computer Engineering, University of Washington, Seattle, USA
[2]Department of Electrical and Computer Engineering, University of Southern California, Los Angeles, USA
[3]Department of Physics, University of Washington, Seattle, USA
[*]Corresponding author: arka@uw.edu

[†]These authors contributed equally

Number of pages: 2
Number of figures: 1

## S1. Ellipsometry measurement of the refractive index of SiN$_x$

We determined the complex refractive index of a 300 nm thick silicon nitride (SiNx) film deposited on silicon using an Angstrom-Sun Technologies Inc. infrared spectroscopic ellipsometer. We collected the Ψ and Δ spectra from 2-15 um and at incidence angles of 65° and 75°. The measured spectra were fitted with a Cauchy dispersion relation for the transparent region and supplemented by two Lorentzian absorption oscillators centered at 10.76 µm and 12.17 µm. This combined model resulted in a 0.9964 goodness of fit across the spectrum using the Levenberg-Marquardt algorithm, provided by the commercial software. The determined complex refractive indices over the 3 – 15 µm wavelength range are presented in Fig. S1.

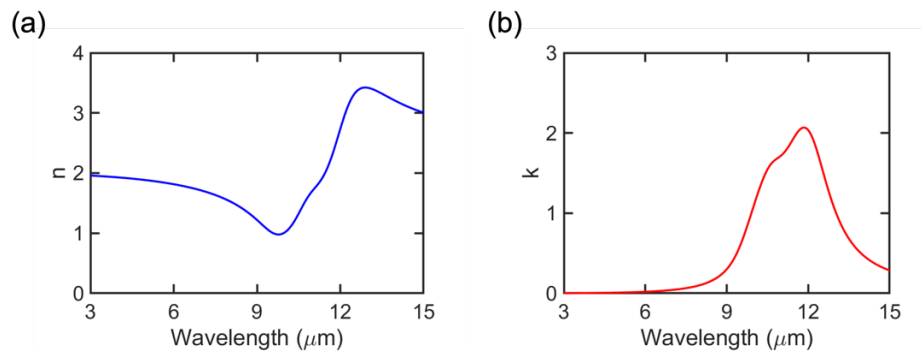

**Figure S1:** Measured refractive index of SiN$_x$.

## S2. Device fabrication

The structure was fabricated on a 500 µm thick Silicon (Si) wafer. Firstly, 15 nm Titanium (Ti) followed by 300 nm Platinum (Pt) back reflector was deposited using electron-beam evaporation (CHA Solution). Following this, 800 nm thick Silicon Nitride (SiN$_x$) was grown using Plasma-Enhanced Chemical Vapor Deposition (PECVD) (SPTS SPM). To make the fractal structure PMMA 495K A6 was spin coated on the sample and exposed using 100kV E-beam lithography (JEOL JBX6300FS). 15 nm Ti followed by 60 nm Pt was next deposited using electron-beam evaporation (CHA SEC-600) and final lift-off of PMMA was performed in Acetone.

## S3. Device characterization

The sample was characterized using Bruker INVENIO-R FTIR and HYPERION 2000 microscope. A 15x objective with NA 0.4 was used to focus light onto the sample. A metallic 0.3 mm confocal aperture was used to get a spot size of 20 µm for the purpose of illuminating only a selected portion of the sample. Each measurement consisted of 64 scans and the resulting data was normalized using a reflection spectrum from a gold mirror.